\input{aipcheck}

\documentclass[
    ,final          
  ]
  {aipproc}

\layoutstyle{8x11double}

\begin{document}

\title{Numerical analysis of impact processes of granular jets}


\author{Tomohiko G. Sano and Hisao Hayakawa}{
  address={Yukawa Institute for Theoretical Physics, Kyoto University Kitashirakawa Oiwakecho, Sakyo-ku, Kyoto 606-8502 Japan}
}

\begin{abstract}
The rheology of a three-dimensional granular jet during an impact is investigated numerically. The cone-like scattering pattern and the sheet-like pattern observed in an experiment [X. Cheng, et al. Phys. Rev. Lett. {\bf 99}, 188001 (2007)] can be reproduced through our calculation. We discuss the constitutive equation for granular jet impact in terms of our simulation. 
From the analysis of an effective friction constant, which is the ratio between the shear stress and the pressure
 the assumption of the zero yield stress would be natural in our setup and the shear visocity is not small in contrast to the suggestion by the experiment.
\end{abstract}

\maketitle


\section{Introduction}

Impact processes of granular flow can be found in wide length scale, not only as problems of natural science but also as those of industrial applications\cite{experiment, inkjet, gsonic1, gsonic2, gsonic3,crater1, crater2, crater3}. A familiar application would be an ink-jet printing, which is an impact process of cohesive grains whose size are within nano-scale\cite{inkjet}. 
Recent experimental and theoretical studies revealed interesting aspects of the impact processes of the granular flow. 
The impact of a granular flow onto a wall, produces a shock, which quantitatively agrees with the Mach cone produced by supersonic gas flow, at low volume fraction\cite{gsonic1, gsonic2, gsonic3}. The impact dynamics of grains is also important as the geophysical problems such as the formation of craters\cite{crater1, crater2, crater3}. One of the recent interesting topics for the impact of the granular flow would be the correspondence between granular flow and Quark Gluon Plasma(QGP), which is expected to behave like the fluid with very small shear viscosity, has been reported experimentally\cite{experiment}. 

Recently, we report that the shear viscosity during the impact is not anomalous, although small shear stress observed in the experiment is reproduced through three-dimensional (3D) simulation. Thus, the correspondence between granular flow and QGP would be superficial\cite{sano_hayakawa}. However, through this analysis, zero yield stress, which is the residual stress without deformation, is assumed, though we presented three indirect evidences to support the assumption. 
In general, this assumption is a strong one, in particular, for the frictional case. 

In this paper, by introducing the effective friction coefficient and the inertia number, which are conventionally used for the dense granular flow\cite{GDR, hatano}, the rheology of the granular jet is discussed. 
Here, we report that the shear viscosity consistent with our previous results, except its density dependence, and the assumption of the zero yield stress would be natural, by performing three-dimensional simulation\cite{sano_hayakawa}.

\begin{figure}
  \includegraphics[scale = 0.80]{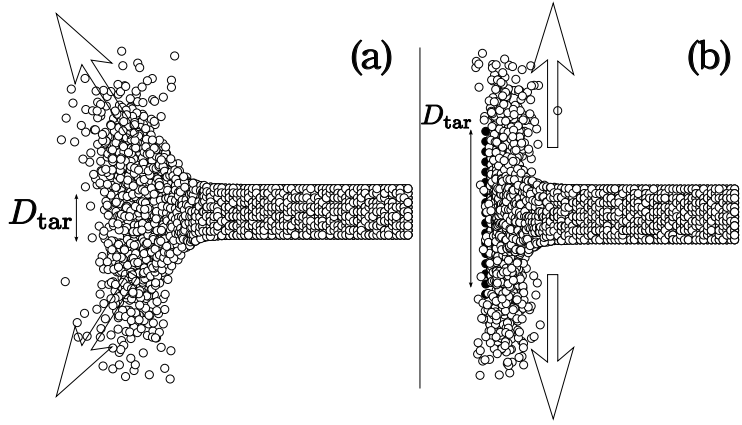}
  \caption{The side view of snapshots for the three dimensional simulation. For $D_{\rm tar} / D_{\rm jet} = 0.8$, cone-like scattered jet is reproduced (a), where black wall particles are hidden by scattered jet. For $D_{\rm tar} / D_{\rm jet} = 2.5$, sheet-like scattered jet is reproduced (b).}
  \label{snapshot}
\end{figure}

\section{Model}

We adopt 
the discrete element method (DEM) for mono-disperse soft core particles of the diameter $d$\cite{DEM}.
When the particle $i$ at the position ${\bf r}_i$ and the particle $j$ at ${\bf r}_j$ are in contact, 
the normal force $F^n _{ij}$ is described as $F^n _{ij}  \equiv  F_{ ij} ^{\rm (el)} + F_{ ij} ^{\rm (vis)}$ with  $F_{ ij} ^{\rm (el)}  \equiv  k_n(d-r_{ ij})$ and $F_{ ij} ^{\rm (vis)}  \equiv -\eta_n({\bf g}_{ ij} \cdot \hat{\bf r}_{ ij})$,
 where  $r_{ij}  \equiv  |{\bf r}_{i} - {\bf r}_{j}|$ and  ${\bf g}_{ij}  \equiv  {\bf v}_{ i} - {\bf v}_{ j}$ 
with the velocity ${\bf v}_i$ of the particle $i$.
 The tangential force is given by $F^t _{ij} \equiv \min\{ |\tilde{F^t _{ij}}|, \mu F_{ij}^n \}{\rm sgn}(\tilde{F}_{ij}^t) $, where  $\mu$ is the local friction constant between contacting grains,
 ${\rm sgn}(x)=1$ for $x\ge 0$ and ${\rm sgn}(x)=-1$ for otherwise, 
$\tilde{F^t _{ij}} \equiv k_{ t} \delta^t _{ij} - \eta_{ t} \dot{\delta}_{ij} ^t$
with the tangential overlap $\delta^t _{ij}$ between $i$ and $j$ particles and the tangential component of relative velocity $\dot{\delta}^t_{ij}$ between $i$ th and $j$ th particles. Here, we adopt parameters $k_t = 0.2 k_n, \eta_t = 0.5 \eta_n$, $\mu = 0.2$, $k_n = 4.98 \times 10^2 mu_0 ^2 /d^2$, $\eta_n = 2.88 u_0 /d$ and $\mu = 0.2$, with incident velocity $u_0$ and the particle mass $m$. 
This set of parameters implies that the restitution coefficient for normal impact is ${e} = 0.75$ and duration time is $t_c = 0.10 d/u_0$. 
The value of $\mu$ is close to the experimental value for nylon spheres.\cite{Rosato}
We adopt the second-order Adams-Bashforth method for the time integration with the time interval $\Delta t = 0.02 t_c$. 

Initial configurations are generated as follows: We prepare fcc crystals and remove particles randomly to reach the desired density. 
We control the initial volume fraction $\phi_0 / \phi_{\rm fcc} \equiv \tilde{\phi_0}$ before the impact as $0.30 \leq \tilde{\phi_0} \leq 0.90$ with volume fraction for a fcc crystal $\phi_{\rm fcc} \simeq 0.74$ and 20,000 particles are used. 
The initial granular temperature, which represents the fluctuation of particle motion, is zero. 
The wall consists of one-layer of particles, which are connected to each other and with their own initial positions via the spring and the dashpot with spring constant $k_p = 10.0 mu_0 ^2 /d^2$ and the dashpot constant  $\eta_p = 5.0 \eta_n$, respectively. 

Experimentally, it is known that the scattered state exhibits the crossover from a cone-like pattern and a sheet-like pattern by changing $D_{\rm tar}/D_{\rm jet}$ with the jet diameter $D_{\rm jet}$ and the target diameter $D_{\rm tar}$. The crossover can be reproduced through DEM, where the jet diameter $D_{\rm jet}$ is fixed $D_{\rm jet} /d = 4.5$ (Fig. \ref{snapshot}). 
White particles (open circles in Fig.1) are grains and black solid ones are wall particles. The Figure \ref{snapshot} (a) is a typical con-like pattern with $D_{\rm tar} / D_{\rm jet} = 0.8$. The Figure \ref{snapshot} (b) is an example of the sheet-like pattern with $D_{\rm tar} /D_{\rm jet} = 2.5$ (b). We note that wall particles are hidden in (a).

\section{Rheology of Granular Jets}
We evaluate physical quantities near the wall at the height $z = \Delta z = 5.0 d$ from the wall $z=0$. $D_{\rm jet} /d = 10.0$ and $D_{\rm tar} /d = 22.0$ .
We adopt the cylidrical coordinate whose symmetric axis is chosen to be the jet axis, 
and divide calculation region into the radial direction $r = 0, \Delta r, \cdots, 5\Delta r$, with $\Delta r \equiv R_{\rm tar} /5$ with the target radius $R_{\rm tar}$.
Then we estimate physical quantities in the corresponding mesh region with $k\Delta r < r < (k+1)\Delta r$ ($k = 0,1,\cdots,5$), where $r$ is the distance from the symmetric axis of the cylindrical coordinate.

We calculate stress tensor as in Ref. \cite{stresstensor}. The microscopic definition of the stress tensor at {\bf r} is given by
\begin{equation}
\sigma_{\alpha \beta}({\bf r}) = \frac{1}{V} \sum_{i} m u_{i \alpha} u_{i \beta} + \frac{1}{V} \sum_{i<j} F_{\alpha} ^{ij} r_{\beta} ^{ij},
\end{equation}
where $i$ and $j$ are indices of particles, $\alpha, \beta = r,\theta,z$ denotes cylindrical coordinates and $\sum_i$ denotes the summation over the particles denoted by $i$ located at ${\bf r}$. 
Here, $z$ axis is parallel to the incident jet axis, and 
$V$ is the volume of each mesh at ${\bf r}$ and $u_{i \alpha} ({\bf r})= v_{\alpha} ^{i} - {\bar v}_{\alpha}({\bf r})$ with the mean velocity $ {\bar v}_{\alpha}({\bf r})$ in the mesh at ${\bf r}$.
To calculate the stress tensor in cylindrical coordinates, we firstly calculate $\sigma_{\alpha' \beta'}$ in Cartesian coordinate, $\alpha', \beta' = x,y,z$, whose origin is the same as cylindrical one, and transform it into that for cylindrical one.

\subsection{Velocity profile}
\begin{figure}
  \includegraphics[scale = 0.60]{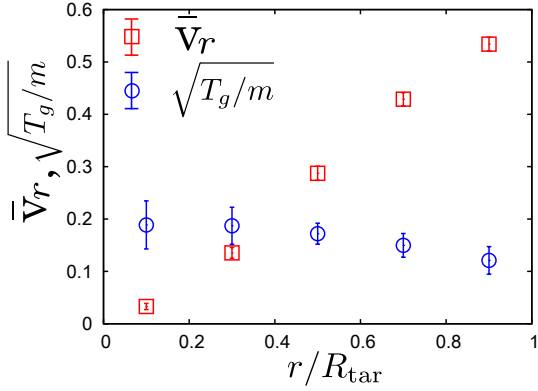}
  \caption{The profile of the radial component of the velocity field $\bar{v}_r$ and $\sqrt{T_g /m}$.}
  \label{vel_temp}
\end{figure}

The profile of $\bar{v}_r$ and $\sqrt{T_g /m}$ is shown in Fig. \ref{vel_temp}, with the granular temperature $T_g({\bf r}) \equiv \sum_{i \alpha} m u^{2} _{i \alpha}({\bf r}) /3N$. Ellowitz {\itshape et al.} suggests that the dead zone, where the motion of grains is frozen, exist near the target in two dimension (2D) \cite{guttenberg, wendy}. However, as is shown in Fig. \ref{vel_temp},  although the velocity of grains at the center is small, the fluctuation of the particle velocity $\sqrt{T_g /m}$ is the largest at the center. Thus, the motion of particles near the target in 3D is not frozen.
Namely, there is no dead zone in 3D granular jets.
 It should be noted that, in our 2D calculation, $T_g$ is small at the center, i.e. the dead zone actually exists, which will be reported elsewhere.

\subsection{Pressure}
Following hydrodynamical model for the granular flow proposed by Garz\'{o} and Dufty\cite{garzo}, pressure $P \equiv \sum_{\alpha} \sigma_{\alpha \alpha} /3$ is conventionally given by
\begin{equation}
\frac{P}{nT_g} = 1 + 2\phi(1+e)\chi, \label{pressure} \\
\end{equation}
\begin{equation}
\chi= \left\{
\begin{array}{ll}
\frac{1- \phi /2}{(1-\phi)^3} & \mbox{ ($0 < \phi < \phi_f $) } \\
 \frac{(1- \phi_f /2)(\phi_c - \phi_f)}{(1-\phi_f)^3(\phi_c - \phi)} & \mbox{ ($\phi_f < \phi < \phi_c $),}
\end{array}
\right. 
\end{equation}
where $\phi_f = 0.49$, $\phi_c = 0.64$, number density $n$, volume fraction $\phi = n\pi d^3 /6$ and radial distribution function $\chi$\cite{torquato}.

For the frictional case, in general, five equations for rotational degree of freedom are necessary, in addition to those for the translational one.
However, ten equations for frictional grains can be reduced to five equations by introducing effective restitution coefficient $\bar{e}$, if the friction constant $\mu$ is small \cite{jenkins,yoon,saitoh}. According to this simplification we use the effective restitution coefficient $\bar{e} = 0.616$ for ${e}=0.75$ and $\mu = 0.2$, for frictional case in the following analysis.

In our setup, the empirical relation (\ref{pressure}) gives a good approximation for $\phi < \phi_f$, 
while the deviation between numerical data and theoretical curve exists for denser regions near the symmetric axis  
which may result from the singularity $r \simeq 0$ of the cylindrical coordinate. 
For later analysis, thus, we adopt the equation of state $P(\phi, T_g) = nT_g \{ 1 + 2\phi (1 + e) \chi \}$ \cite{sano_hayakawa}.

\subsection{Friction coefficient}
Let us analyze the effective friction coefficient for macroscopic motion of a collection of grains, following the ref. \cite{cruz}. We estimate strain rate $D_{rz}$ as $\partial \bar{v}_r (r,\Delta z/2) / \partial z \simeq ( \bar{v}_r(r,3\Delta z /4) - \bar{v}_r(r,\Delta z /4)) / (\Delta z /2)$ and  $\partial \bar{v}_z (r,z) / \partial r \simeq ( \bar{v}_z(r + \Delta r/2,z) - \bar{v}_z(r - \Delta r/2,z)) / \Delta r$. Since we evaluate the physical quantities near the wall, the mesh $0 < z < \Delta z$ is divided into $0 < z < \Delta z /2$ and $\Delta z /2< z < \Delta z$ to calculate $\partial \bar{v}_r (r,\Delta z/2) / \partial z$ and $0<r<R_{\rm tar}$ is divided into $0<r<\Delta r/2, \Delta r/2<r< 3\Delta r/2, \cdots$. 

\begin{figure}
  \includegraphics[scale = 0.60]{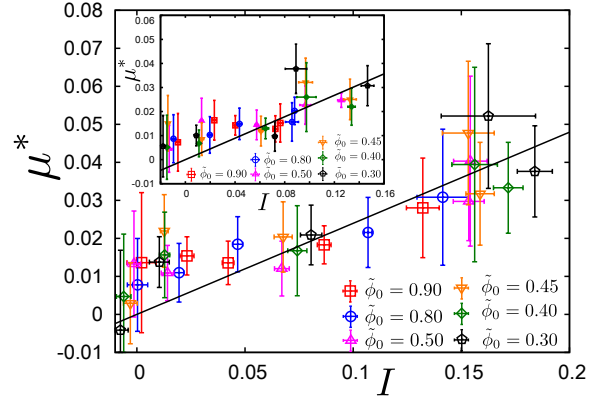}
  \caption{Analysis on the $\mu^*$ vs $I$ plane. Numerical datum can be fitted into black solid lines $\mu^* = a I$.}
  \label{friction}
\end{figure}

Introducing the effective friction coefficient $\mu^* \equiv - \sigma_{rz} / P$ and the inertia number $I \equiv D_{rz} \sqrt{P/md}$, we plot the observed data $\mu^*$ vs $I$ for several $\tilde{\phi}_0$ in Fig. \ref{friction}. From the result of our simulation, the obtained effective friction coefficient $\mu^*$ can be fitted by $\mu^* = a I$ with a constant $a$ within error bars, where fitting values are $a = 0.240$ and $a = 0.223$ for frictional and frictionless case, respectively. The solid lines represent the corresponding fitting lines. Judging from the fitting, the assumption of zero yield stress in our setup\cite{sano_hayakawa} would be natural.

Through this analysis, the constitutive equations are obtained as
\begin{eqnarray}
\sigma_{rz} &=& - \eta D_{rz}\\ 
\eta^* &=&  \frac{16a}{5}\sqrt{6\phi (1+2\phi(1+e)\chi)}.
\end{eqnarray}
Here, non-dimensional shear viscosity $\eta^* = \eta^* (\phi)$ is introduced as $\eta^* \equiv \eta / \eta_0$ with the shear viscosity $\eta$ and $\eta_0 \equiv 5\sqrt{\pi m T_g}/16d^2$ by convention\cite{garzo}. 

\begin{figure}
  \includegraphics[scale = 0.60]{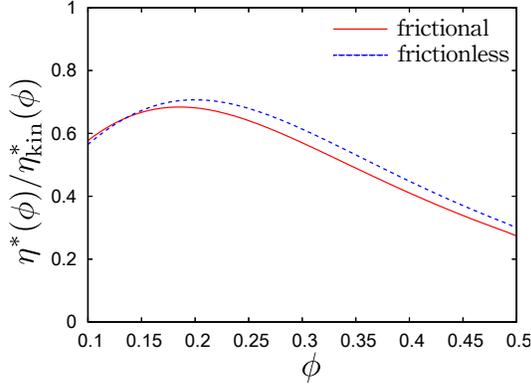}
  \caption{Comparison of the non-dimensional shear viscosity based on the kinetic theory $\eta^* _{\rm kin}(\phi)$\cite{garzo} with our result, where the solid line and the dashed line denotes  for the frictional case and the frictionless case, respectively.}
  \label{comparison}
\end{figure}

The obtained shear viscosity is proportional to $\sqrt{T_g}$, which is consistent with our previous results\cite{sano_hayakawa}. However, the obtained density dependence of $\eta^*$ is different from ref. \cite{garzo}. The non-dimensional shear viscosity based on the kinetic theory $\eta^* _{\rm kin}(\phi)$ is compared with our result in Fig. \ref{comparison}. The solid line and the dashed line denotes  for the frictional case and the frictionless case, respectively. The obtained shear viscosity is less than $80\%$ of that of the kinetic theory. It should be noted that the obtained shear viscosity is also finite, even through the analysis based on the effective friction coefficient.

\section{Discussion}

The difference of density dependence of non-dimensional shear viscosity would be understood in the followings. In the analysis on $\mu^*$ vs $I$ plane, the density dependence of transport coefficient is assumed to appear only through pressure $P$, i.e.
 \begin{equation}
\eta(P, T_g, \phi) = \eta(P(\phi, T_g)).
\label{assp}
\end{equation}
Thus, once the equation of state is determined, $\eta^*(\phi)$ is uniquely obtained. On the other hand, in ref. \cite{garzo}, because density dependence of $\eta^* _{\rm kin}$ results from the kinetic theory, results are different. 
At present, we could not judge which viscosity is better.
However, we should note that the $\eta \propto \sqrt{T_g}$ can be obtained, even under the assumption of eq. (\ref{assp}).

We comment here that $\mu^* = \mu^* (I)$ in two dimensions cannot be fitted by $\mu^*=a I$.
 Because grains are easily packed through the impact in 2D, dense flow and related frictional phenomena, where $\mu^*$ would be greater by a degree of magnitude than that in 3D, emerges. The existence of the dead zone may cause the difference between 2D and 3D results. The frictional phenomena in 2D will be reported elsewhere.

\section{Conclusion}
We have numerically investigated the constitutive equation for the granular jet impact in 3D, introducing the analysis on $\mu^*$ vs $I$ plane. The dead zone, which is suggested by Ellowitz {\itshape et al}\cite{wendy}, can not be reproduced in our 3D study, although the velocity field at the center is small. Rheological results are consistent with our previous results\cite{sano_hayakawa}, except for the density dependence of the shear viscosity, which results from the assumption in Eq. (\ref{assp}). Judging from the analysis, the assumption of zero yield stress would be natural\cite{sano_hayakawa}.

\begin{theacknowledgments}
 This work is partially supported by 
and the Grant-in-Aid for the Global COE program ``The Next Generation of Physics, Spun from Universality and Emergenceh from MEXT, Japan.
\end{theacknowledgments}



\bibliographystyle{aipproc}   

\bibliography{sample}

\IfFileExists{\jobname.bbl}{}
 {\typeout{}
  \typeout{******************************************}
  \typeout{** Please run "bibtex \jobname" to optain}
  \typeout{** the bibliography and then re-run LaTeX}
  \typeout{** twice to fix the references!}
  \typeout{******************************************}
  \typeout{}
 }

\end{document}